\documentclass[twocolumn]{article}
\usepackage{graphicx}
\usepackage{amssymb,amsmath}
\DeclareGraphicsExtensions{.pdf,.gif,.eps,.ps,.jpg,.bmp,.png}

\begin{document}

\title{Non-Markovian Beables vs.\ Massive Parallelism}
\author{Eric Dennis\thanks{}\\ \emph{Morgan Stanley, New York, NY 10036}}
\date{}

\twocolumn[
\begin{@twocolumnfalse}

\vspace{-.5in}

\maketitle

\renewcommand{\abstractname}{}
\begin{abstract}
A simple dynamical model over a discrete classical state space is presented. In a certain limit, it reduces to one in a class of models subsuming Bell's field-theoretic version of Bohmian mechanics. But it exhibits the massive parallelism native to quantum mechanics only as an emergent phenomenon, in contrast with Bell's and other hidden variable theories. While still non-local in its dynamics, the model thus restores our ability to regard a system as a combination of separate, localized parts, at the price of admitting non-Markovian dynamics.
\end{abstract}
\end{@twocolumnfalse}
]
{
\renewcommand{\thefootnote}%
{\fnsymbol{footnote}}
\footnotetext[1]{ericmdennis@gmail.com}
}

\noindent There are two different kinds of weirdness attributed to quantum mechanics. Starting with Bohr, well prior to the development of quantum theory itself, many have asserted the need to revise certain philosophical propositions associated with the classical world view. The original such proposition was supposed to be classical determinism, but later the target moved to something deeper: the observer-independence of physical reality. A frontal assault on this target flows from the shifty distinction between subject and object, measurer and measured, at the core of orthodox interpretations of quantum mechanics. Hidden variable theories were devised by those (most notably Einstein, David Bohm, and John Bell) who sought to retain, not primarily determinism, but the concept of a mind-independent reality and theories that talk about it.

Another kind of weirdness is brought out by Bell's theorem, which gives the final blow to one cherished premise---locality---in any viable theory, orthodox or otherwise.\footnote{Bell's theorem applies as long as one accepts the validity of measurement data, whose seeming inconsistency with unitary evolution is the source of the interpretation problem. The many worlds interpretation evades this by baldly denying that the specific measurement outcome we observe, singling out one eigenstate, actually occurs.} Opposing what has become a popular view, Bell's own careful examination \cite{Sox}\cite{Travis} of the theorem renders it a charge against locality as such, failing to reveal any other relevant premise (usually going under the rubric of ``realism,'' or sometimes even less coherently ``determinism'') to take the fall. In any case, quantum evolution entails a localized disturbance instantaneously redounding throughout a spatially extended system. Such effects are manifest in current formalisms, whether in terms of a collapsing state vector or a non-local guidance equation. And whether or not these effects happen to be exploitable by man to send rapid signals does not alter the fact of their existence, nor the challenge they pose to Lorentz invariance as a fundamental (rather than emergent) symmetry of nature.

But a more worrisome non-locality is wrapped up in the very concept of the quantum state. The state vector generally cannot be understood as a combination of separate, localized pieces. Rather, insofar as it can be analyzed into pieces at all, they correspond to all the different possibilities for the classical configuration of the whole system. The simultaneous coexistence of all these configurations (``massive parallelism'') in a single quantum state is the real basis for quantum weirdness and potentially the source of an unprecedented computational power ready to be harnessed by man.

No serious hidden variable theory has dispensed with this massive parallelism; it is always present in the ``beables.''\footnote{ This is Bell's term for things that, according to a theory, may actually \emph{be}, as opposed to the ``observables'' posited by quantum theory, which are not supposed to have any mind-independent existence.} In Bohmian mechanics, for instance, even though one classical state is singled out in each moment as physically real, we retain the massively parallel state vector as a separate beable, guiding the evolution of that state \cite{Shelly}. Although a kind of conspiratorial Bohm-like model has been suggested as proof-of-concept for beable theories lacking such an object \cite{TELB}.

I will present a simple, general model that exhibits the massive parallelism of quantum mechanics as an emergent phenomenon. In this model, a variation on Bell's field-theoretic version \cite{Bell} of Bohmian mechanics, the real state of the world at any one moment is a single, classical one. While the dynamics is non-local, the non-locality does not arise (as it does in Bell's model) from a massively parallel object lurking in the background. The whole system may be regarded, again, as a combination of localized, albeit non-locally interacting, physical parts. This is accomplished by surrendering the assumption of Markovian dynamics. Massive parallelism is effectively replaced by the ability of the system to remember massive numbers of its past states.

\section{Defining the model}

Consider a discrete classical state space given by the nodes $n \in N$ of an undirected graph $G$, whose edges $e = \{n,m\}\in E$ then define the allowed transitions between these states. The system state proceeds in time over the trajectory given by a sequence $(n_1,n_2,\ldots)$ of nodes, such that each pair $e_k = \{n_{k-1},n_k\}$ is in $E$, together with the times $(t_1,t_2,\ldots)$ at which the respective nodes are reached. We may set $t_1 = 0$.

Occupying one of these nodes $n_k$ at some time $t$, the system will be capable of jumping over any edge $e$ in the set $E_k \subset E$ of edges touching $n_k$. To each $e=\{n_k,m\}$ in $E_k$ we associate a complex number $A_k(e)$ with units of energy, called a \emph{jump potential}, which determines the probability, $T_k(e)dt$, of the transition $n_k \rightarrow m$ occurring in the interval $(t,t+dt)$:
\begin{equation}\label{jumprate}
T_k(e) = -\mathrm{Im}[A_k(e)]/\hbar + |A_k(e)|/\hbar_2
\end{equation}
where $\hbar_2 \ll \hbar$ is a new constant.

\begin{figure}[ht]
\centering
\includegraphics[scale=0.6, bb = 1 1 300 250]{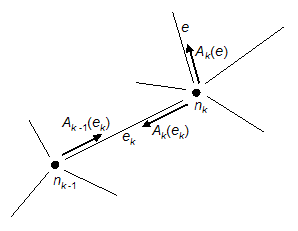}
\caption{\small Nodes, edges, and jump potentials involved in the update rule (\ref{potentials}).}
\label{e_k}
\end{figure}

The jump potentials themselves are determined as follows. Upon the transition $n_{k-1} \rightarrow n_k$, the system remembers the potentials associated with edges $e \in E_k$ from the last time $t_{\dot k}$, before $t_k$, that a transition involving $n_k$ occurred:
\begin{equation}\label{remember}
A_k(e) = A_{\dot k}(e)
\end{equation}
In case no such previous transition occurred, we refer to some set of initial values for the jump potentials that may depend on $n_k$ and $e$. Just after applying (\ref{remember}) at $t_k$, we update the two potentials associated with $e_k$ according to the rules
\begin{eqnarray}\label{potentials}
A_{k-1}(e_k) &\rightarrow& A_{k-1}(e_k) \, e^{-i A_k(e) (t_k-t_{\bar k})/\hbar} \nonumber \\
A_k(e_k) &\rightarrow& A_k(e_k) \, e^{i A_k(e) (t_k-t_{\bar k})/\hbar}
\end{eqnarray}
for each $e \in E_k$, where $t_{\bar k}$ is the last time before $t_k$ that the transition $n_{k-1} \rightarrow n_k$ occurred. In case it has not occurred before, we put $t_{\bar k} = 0$. Note that $A_{\dot k}(e)$ in (\ref{remember}) is understood as the post-update value, after (\ref{potentials}) was invoked at $t_{\dot k}$.

\section{The dynamics as $\hbar_2 \rightarrow 0$}

The dynamics of this model in the limit $\hbar_2 \rightarrow 0$ implies ergodicity over $G$ for any finite time period, which is to say the jump potentials will be changing through (\ref{potentials}) very rapidly. Let us keep track of these changes over time by using the notation $A_{nm}(t)$ for the jump potential corresponding to the transition $n \rightarrow m$, with the trajectory passing through it at each of the times $(t_{k_1}, t_{k_2},\ldots)$, so, \emph{e.g.}, $\bar{k_j} = k_{j-1}$ and $A_{nm}(t) = A_{k_j-1}(e_{k_j})$ for $t \in (t_{k_j-1},t_{k_j})$.

According to (\ref{potentials}), the $n \rightarrow m$ jump occurring at $t_{k_j}$ will have caused $A_{nm}$ to be multiplied by a factor
\[
e^{-iA_{mp}(t_{k_j})(t_{k_j}-t_{k_{j-1}})/\hbar}
\]
for each $p$ such that $\{m,p\} \in E$. These accumulating updates for $A_{nm}$ will produce a total factor $e^{-iS_{mp}/\hbar}$ for each such $p$, with $S_{mp}$ given by
\begin{equation}\label{Smp}
\sum_j A_{mp}(t_{k_j})(t_{k_j}-t_{k_{j-1}}) \rightarrow \int_0^t A_{mp}(t^\prime)dt^\prime
\end{equation}
The passage to an integral occurs as $\hbar_2 \rightarrow 0$, in which case $t_{k_1} \rightarrow 0$. Updates to $A_{nm}$ also occur when the trajectory makes a jump $m \rightarrow n$ so that (\ref{potentials}) involves a factor of the form
\[
e^{iA_{np}(t_{k_j})(t_{k_j}-t_{k_{j-1}})/\hbar}
\]
for each $p$ such that $\{n,p\} \in E$, where the $t_{k_j}$ now refer to times at which these $m \rightarrow n$ jumps occur. This produces an accumulating update factor $e^{iS_{np}/\hbar}$ similar to the one above. The total result of (\ref{potentials}) is an evolving jump potential given by
\begin{equation}\label{A}
A_{nm}(t) = A_{nm}(0) \frac{ e^{ -i \sum_p \int^t_0 A_{mp}(t^\prime) dt^\prime/\hbar } }
                                                { e^{ -i \sum_p \int^t_0 A_{np}(t^\prime) dt^\prime/\hbar } }
\end{equation}
as $\hbar_2 \rightarrow 0$, where we understand that $A_{np} \equiv 0$ if $\{n,p\} \notin E$.  In differential form this becomes
\begin{equation}\label{dAdt}
i \hbar \frac{dA_{nm}}{dt} =  A_{nm} \sum_p ( A_{mp}-A_{np} )
\end{equation}

\section{Recovering QM}

We can now begin to make the connection to quantum mechanics. The form of (\ref{A}) allows us to write, in the $\hbar_2 \rightarrow 0$ limit,
\begin{equation}\label{Hpsi}
A_{nm}(t) = H_{nm} \frac{ \psi_m(t) }{ \psi_n(t) }
\end{equation}
for some new, formal quantities: a time-independent $H$ and time-dependent $\psi(t)$. In particular, let us take (\ref{Hpsi}) at $t=0$ as defining the initial values $A_{nm}(0)$. The jump rates (\ref{jumprate}) in this limit formally resemble those of Bell \cite{Bell} and constitute a degenerate member of the class of jump rates described in \cite{Bac}. If we identify the denominator on the right side of (\ref{A}) with $\psi_n(t)/\psi_n(0)$, it follows immediately that
\begin{equation}\label{schrod}
i \hbar \frac{d\psi_n}{dt} = \sum_p H_{np} \psi_p
\end{equation}
recognizable as the Schrodinger equation over our discrete state space. Thus the evolution of the jump potentials along an ergodic trajectory in the $\hbar_2 \rightarrow 0$ limit fully encodes the evolution of the conventional state vector $\psi$.

Assuming that $H$ is hermitian, we have $A^\ast_{nm} = A_{mn}|\psi_m|^2/|\psi_n|^2$ from (\ref{Hpsi}), which, along with (\ref{jumprate}), implies
\[
T_{mn}|\psi_m|^2 - T_{nm}|\psi_n|^2 = 2 \mathrm{Im}(A_{nm}) |\psi_n|^2/\hbar
\]
independently of $\hbar_2$, using an obvious notation for the transition rates. This together with (\ref{Hpsi}) and (\ref{schrod}) gives
\begin{equation}\label{dpsi}
\frac{d}{dt}|\psi_n|^2 = \sum_m ( T_{mn} |\psi_m|^2  - T_{nm} |\psi_n|^2 )
\end{equation}
If we consider an ensemble of trajectories evolving independently (and carrying independent sets of jump potentials) according to (\ref{jumprate}) and (\ref{potentials}), we see that (\ref{dpsi}) is exactly the equation obeyed by the statistical density $\rho_n(t)$ of trajectory endpoints over the state space. Therefore, if $\rho_n = |\psi_n|^2$ holds at $t=0$, it will hold at all subsequent times.

The ensemble statistics in the $\hbar_2 \rightarrow 0$ limit will mirror $|\psi(t)|^2$ provided the initial values $A_{nm}(0)$ are set according to (\ref{Hpsi}) at $t=0$, with $H$ identified as the system Hamiltonian, $\psi(0)$ as the initial state vector, and $E$ as the set of pairs $\{n,m\}$ such that $H_{nm} \ne 0$. (Note that this will generally imply the presence of looped edges $\{n,n\}$ so that sometimes $n_k = n_{k-1}$, $A_k = A_{k-1}$, and the two rules (\ref{potentials}) will cancel each other, producing a trivial update at $t_k$.)

Interestingly, the evolution laws (\ref{jumprate}) and (\ref{potentials}), as opposed to (\ref{schrod}), are non-parametric (except for $\hbar$ and $\hbar_2$). Apart from the state space topology encoded in $E$, the complexity of the Hamiltonian no longer resides in the dynamics but rather in the initial conditions. This is not to assert the uniqueness of (\ref{jumprate}), whose dependence on $A_k(e)$ may be altered without affecting (\ref{dpsi}), or of (\ref{potentials}), which may be replaced by rules involving different edges without affecting (\ref{dAdt}).

A technical problem might arise with the possibility of $\psi_n(t)=0$ for some $n$ and $t$. In this case $A_{nm}(t)$ ought to diverge and $A_{mn}(t)$ vanish at $t$, but then (\ref{potentials}) seems to imply this must persist for all time after $t$. This does not appear to be a practical impediment as we may (in the $\hbar_2 \rightarrow 0$ limit) set initial values for the jump potentials arbitrarily small or large, but finite, and (\ref{potentials}) will ensure finiteness going forward.

\section{Experimental signatures}

The Schrodinger equation (\ref{schrod}) is in general recovered when $\hbar_2$ is small enough that the system trajectory rasters over the entire state space much faster than the $A_{nm}$ are appreciably changing. Even then, over sufficiently long time scales, the discreteness effects of the sum (\ref{Smp}) might build up to stear $A_{nm}$ away from the solution of (\ref{dAdt}). Moreover, we may still express $A_{nm}$ according to (\ref{Hpsi}), but now $\psi(t)$ will evolve by some very complicated discretized version of (\ref{schrod}), with the discreteness effects contributing something like an extra noise term including non-linear, non-Markovian feedback from (\ref{jumprate}). The result will be a departure from unitary evolution, hence a sub-quantum effect that will, however, look like decoherence, even absent environmental degrees of freedom.

The leading contribution to deviations of the integral from the sum in (\ref{Smp}) is
\[
\sum_j {\textstyle\frac{1}{2}} A^\prime(t_{k_j}) (t_{k_j}-t_{k_{j-1}})^2 \sim \hbar \omega t_\mathrm{rec}
\]
where $\omega^{-1}$ is a timescale associated with $H$, and $t_\mathrm{rec} \sim t_k - t_{\bar k}$ is a typical expected recurrence time, which is controlled by the $\hbar_2$ term in (\ref{jumprate}) and should be roughly given by
\begin{equation}\label{trec}
t_\mathrm{rec} \sim \frac{\hbar_2}{\hbar \omega} |N|^\gamma
\end{equation}
if $\psi$ exhibits massive parallelism, where $|N|$ is the total number of system states. The exponent $\gamma = 1$ applies for homogeneous diffusive behavior in a high dimensional state space, which we will assume; it may need to be modified depending on the dynamics for a particular system. If these deviations are not observed, we get the crude bound
\begin{equation}\label{bound}
\hbar_2 \lesssim \hbar/|N|
\end{equation}

Experimentally, many-body systems demonstrating quantum coherence for massively parallel states would seem of interest here, as $|N|$ will scale exponentially with the number of degrees of freedom. But it may be possible to generate such effects by maintaining a superposition of just two states far-separated in state space, \emph{i.e.}\ a macro- or mesoscopic superposition, if one is careful to exclude any preferred pathway between them. Otherwise $|N|$ in (\ref{bound}) may be effectively reduced to the number of states along such a pathway, which could be $\sim \log |N|$. A more credible bound would require using the model to simulate a particular system and get some idea about the scaling of its sub-quantum `decoherence' time $t_\mathrm{dec}(\hbar_2,|N|)$.

Additionally, if such a system could be probed on the timescale of $t_\mathrm{rec}$, one would expect to observe the failure of the realized state $n_k$ to fully sample state space and thus disseminate (\ref{potentials}). In particular consider preparing a ``cat state'' $(|0\cdots 0\rangle + |1\cdots 1\rangle)/\sqrt{2}$ of length $N_\mathrm{qubits}$: prepare $|0\cdots0\rangle$, rotate the first qubit $|0\rangle \rightarrow (|0\rangle + |1\rangle)/\sqrt{2}$, and perform $N_\mathrm{qubits}-1$ controlled-not gates from the first qubit into all of the others. Finally, measure $\sigma^i_z$ on each qubit $i$. 

Assuming reasonable parallelization, the time taken to perform all these gates and measurements will be roughly $1/\omega$, since $H = H_\mathrm{sys}+H_\mathrm{int}$ (see the next section) will be dominated by $H_\mathrm{int}$. If $t_\mathrm{rec} \ll 1/\omega$, then $n_k$ will have trouble ever reaching the state $|1\cdots 1\rangle$ from its initial location at $|0\cdots 0\rangle$. With $n_k$ stuck around $|0\cdots 0\rangle$, the $\sigma_z$ measurements will each yield $+1$ with high probability; whereas the quantum result is $\sigma_z = \pm 1$ with equal probability. Define the ``total spin'' $M = \sum_i \langle \sigma^i_z \rangle$. We would expect to reproduce the quantum result $M = 0$ for $t_\mathrm{rec} \ll 1/\omega$ but find that $M \rightarrow N_\mathrm{qubits}$ if $t_\mathrm{rec} \gg 1/\omega$. Failing to observe appreciable $M$ values would again yield the crude bound (\ref{bound}), which again ought to be tested by means of detailed simulation for particular systems. 

\section{Measurement theory}

As in Bell's model, one would like here as well to have a Bohmian, \emph{i.e.} observer-independent, account of the physical processes of measurement. One must be careful, though, because two limits are involved in such an account: the $\hbar_2 \rightarrow 0$ limit and the decoherence limit originally conceived by Bohm \cite{Bohm} to explain the practically irreversible entrapment of the system trajectory within the particular region of (joint system-apparatus) state space corresponding to a distinct measurement result. The decoherence limit entails a large number $d_\mathrm{env}$ of degrees of freedom in the measuring apparatus becoming entangled with the system.

It is perhaps easiest to think in terms of an effective description of the system-apparatus interaction as time-dependent. We can incorporate this into the model by adding another rule to (\ref{potentials}):
\[
A_k(e_k) \rightarrow A_k(e_k) \frac{H_k}{H_{\bar k}}
\]
where we regard the time-dependent Hamiltonian $H_k \equiv H_{n_k n_{k-1}}(t_k)$ as information accessible at $n_k$. This modifies the right side of (\ref{A}) by introducing an extra factor of $H_{nm}(t)/H_{nm}(0)$ and turns (\ref{dAdt}) into
\[
i \hbar \frac{dA_{nm}}{dt} =  A_{nm} \left[\sum_p ( A_{mp}-A_{np} ) + i \frac{d}{dt}\log H_{nm} \right]
\]
We recover the usual Schrodinger equation with time-dependent $H$ by an argument paralleling the previous one and can now contemplate arbitrary unitary (as $\hbar_2 \rightarrow 0$) entangling operations between system and apparatus.

Each old node $n$ is identified with a product state between the system in $n$ and the apparatus in some fixed initial state. Turning on the system-apparatus interaction means turning on edges connecting the old nodes with new ones representing product states involving different configurations of the apparatus. 

Holding $d_\mathrm{env}$ constant and taking $\hbar_2 \rightarrow 0$ will produce a system trajectory in the joint space that tracks the joint $\psi(t)$ in its entirety for all time, which is to say it will utterly fail to explain the observed reality of a particular system eigenstate being singled out by measurement. The result is something akin to the many worlds intepretation: the trajectory is rastering over state space so quickly and so ergodically that the notion of a single state $n$ being occupied at a given moment dissolves---not even decoherence can effectively confine it to a particular region.

However, holding $\hbar_2$ well bellow $\hbar$ but fixed as we increase $d_\mathrm{env}$ gives a different result, dependent on the structure of the joint state space. Suppose we can gain some information about the system by allowing it to interact with a ``pointer'' degree of freedom $P$ in the apparatus, which will discriminate between some system states $|\phi_n\rangle$ by entangling with them according to a unitary transformation acting as
\begin{equation}\label{pointer}
|\phi_n\rangle |P_0\rangle \rightarrow |\phi_n\rangle |P_n\rangle
\end{equation}
where the $|P_n\rangle$ are orthogonal states of the pointer. Unitarity then requires that the $|\phi_n\rangle$ be orthogonal among themselves. (We are considering the non-degenerate case, but the degenerate one is a simple extension.) Thus the $|\phi_n\rangle$ define a Hermitian operator that has them as its eigenstates. The system-apparatus coupling (\ref{pointer}) is described as (the first step in) a ``measurement'' of this operator.

For a state corresponding to some $n \in N$, (\ref{pointer}) implies the action
\begin{equation}\label{pointer2}
|n\rangle |P_0\rangle \rightarrow \sum_{n^\prime m} \langle\phi_m|n\rangle \langle n^\prime|\phi_m\rangle |n^\prime\rangle|P_m\rangle
\end{equation}
which may be translated into an effective Hamiltonian $H = H_\mathrm{sys} + H_\mathrm{int}(t)$ linking new joint nodes $(n^\prime,P_m)$ to the old nodes $n$, identified respectively with $(n,P_0)$. We now have a system-pointer joint space of size $|N|^2$. $P$ will proceed to imprint itself on additional degrees of freedom in the apparatus, which will themselves do likewise, in a cascading process of decoherence. 

\begin{figure}[ht]
\centering
\includegraphics[scale=0.7, bb = 1 1 260 280]{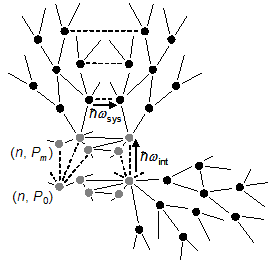}
\caption{\small Decoherence pictured in terms of sub-networks of new states spawned in trees from system-pointer joint states $(n,P_m)$, shown as grey nodes, themselves arising ($H_\mathrm{int} \sim \hbar \omega_\mathrm{int}$) from the pre-measurement states $(n,P_0)$. Trees spawned from nodes with the same $P_m$ are isomorphic and inherit inter-tree links from $H_\mathrm{sys} \sim \hbar \omega_\mathrm{sys}$; trees associated with distinct $P_m$ are disconnected.}
\label{decoherence}
\end{figure}

Each joint node $(n,P_m)$ from the original pointer interaction will spawn a tree of additional nodes connected by couplings within the apparatus. Since the additional degrees of freedom couple only to $P$, the two trees spawned by $(n,P_m)$ and $(n^\prime,P_m)$ respectively will be isomorphic in terms of their branch structure and Hamiltonian matrix elements. Indeed corresponding nodes in these two trees will be connected exactly when $(n,P_m)$ and $(n^\prime,P_m)$ are connected. On the other hand, the two trees spawned by $(n,P_m)$ and $(n^\prime,P_{m^\prime})$ with $m \ne m^\prime$ will not link to each other as we proceed further into their branch structures, since the respective apparatus states will have diverged over many different degrees of freedom. (See Fig.\ \ref{decoherence}.) The trees will thus grow into a set of diverging sub-networks corresponding to the different pointer states $P_m$, connecting with each other only around the original joint nodes $(n,P_m)$.

Right before the system-apparatus interaction is turned on, $\hbar_2 \ll \hbar$ implies that $\rho_n = |\langle n|\psi \rangle|^2$ holds to high accuracy (as a statistical distribution over an ensemble of realized states $n_k$). As the system couples to $P$, the action on the quantum state $|\psi\rangle$ of the system is given by (\ref{pointer}) as
\[
|\psi\rangle|P_0\rangle \rightarrow \sum_m \langle\phi_m|\psi\rangle |\phi_m\rangle|P_m\rangle
\]
so $n_k$ will wind up somewhere inside the set $\{(n,P_m)|n \in N\}$ with probability $|\langle \phi_m | \psi \rangle|^2$. But one must remember that this is an effective description of the fundamental dynamics (\ref{jumprate}) and (\ref{potentials}). It will hold if $t_\mathrm{rec} \lesssim 1/\omega_\mathrm{int}$ for the pointer-system state space so that (\ref{potentials}) is adequately disseminated across this space before decoherence begins in earnest. Assuming that $\omega_\mathrm{int}$ characterizes all the apparatus couplings and that $\omega_\mathrm{int} \gg \omega_\mathrm{sys}$, this occurs when
\[
d_\mathrm{sys} \lesssim \log ( \hbar / \hbar_2 )
\]
where $d_\mathrm{sys} \sim \log |N|$.

As new trees are spawned, $n_k$ (now understood as occupying an arbitrary node in the full joint space) will diffuse into its associated sub-network. Ensemble-wise, $\rho$ will diffuse into all of the sub-networks. With the sub-network sizes increasing exponentially in $d_\mathrm{env}$, the probability that $n_k$ ventures all the way back to the beginning of one of these trees in order to move into a different sub-network is exponentially suppressed. The threshold of effective irreversibility is passed when the system-apparatus state space is large enough that $t_\mathrm{rec} \gtrsim 1/\omega_\mathrm{int}$, hence
\[
d_\mathrm{env} \gtrsim \log ( \hbar / \hbar_2)
\] 

Pursuing a remark made above, one could even extend the present formalism to provide a continuous bridge between many worlds and Bohmian mechanics by replacing (\ref{potentials}) with rules updating all edges within a distance $\Delta n$ of $n_k$. The $\hbar_2 \rightarrow 0$ limit yields many worlds, the $(\hbar_2,\Delta n) \rightarrow (\hbar,|N|)$ limit yields what is morally Bohmian mechanics, and the $\Delta n \rightarrow 0$ limit at fixed $\hbar_2 \ll \hbar$ is the regime advocated here. 

As opposed to Bohmian mechanics, once the realized state has become effectively confined to one decohered region (sub-network) of state space in the present model, not only do the $A_{nm}$ from other regions cease to influence the trajectory, but they cease to receive any updates via (\ref{potentials}) at all---they are forgotten and \emph{frozen} into history. Furthermore, it is no longer just the brute size of $d_\mathrm{env}$, but now also a new scale determined by $\hbar/\hbar_2$ that provides an objective basis for the micro/macro distinction and the emergence of the classical limit.

\section{Discussion}

Despite the identification (\ref{Hpsi}) and the formal resemblance of the derivation of (\ref{dpsi}) here to that of Bell \cite{Bell}, the present model is not to be understood as describing the motion of a classical system guided by an evolving state vector, in the style of Bohmian mechanics. The evolution of such a massively parallel state vector occurs simultaneously at every point in state space, suggesting the simultaneous reality of all possible classical states.

In the present model, however, the only beables at a given time, the only things that exist, are the one classical state $n_k$ and the jump potentials $A_k(e)$ associated with allowed transitions out of $n_k$. There are no beables referring to states far away from $n_k$ in state space, hence there is fundamentally no massive parallelism. We can maintain this because, unlike the conventional state vector, a given jump potential is updated by (\ref{potentials}) only when the realized state is actually undergoing a transition associated with that potential.

This kind of locality in the evolution of a single trajectory over state space should not be mistaken for locality in physical space. Physical non-locality in the model derives from the fact that the probability to undergo a particular transition $n \rightarrow m$ depends on the state of the whole system even if $m$ only differs from $n$ by one local operation. To put it more formally, suppose $X$ and $Y$ are operations localized around two distant physical positions $x$ and $y$ respectively, and suppose
\begin{eqnarray*}
|m\rangle &=& Y|n\rangle \\
|n^\prime\rangle &=& X|n\rangle \\
|m^\prime\rangle &=& X|m\rangle
\end{eqnarray*}
all correspond to states in $N$. Typically $A_{nm} \neq A_{n^\prime m^\prime}$, and similarly for the transition probabilities. Thus the model permits something at $x$ to instantaneously influence a process occurring at $y$.

One might object to the claim that only $n_k$ and $A_k(e)$ exist at a given time, since in order to perform successive updates according to (\ref{potentials}) the system effectively needs to keep a record of past values of these beables for as far back as $t_{\bar k}$. It would be more compelling to eliminate massive parallelism rather than exchange it for a radically non-Markovian dynamics, but one suspects doing so would be tantamount to supplying a general algorithm for efficient classical simulation of quantum systems. Indeed, for quantum dynamics problems over a finite dimensional Hilbert space, the present model offers a one-parameter family of simulations, indexed by $\hbar_2$, that will converge to the real quantum evolution as $\hbar_2 \rightarrow 0$. But the utility of such an algorithm will depend in particular cases on the behavior of the function $t_\mathrm{dec}(\hbar_2,|N|)$ described above.

Conversely, if a model such as this one were to actually lie underneath quantum mechanics, massive parallelism would not be an arbitrarily scalable resource, and the viability of quantum information as a technology might, in principle, be affected. Scalability would be limited by $N_\mathrm{qubits} \lesssim \log(\hbar/\hbar_2)$. 

On an even more speculative note, the strongly non-Markovian nature of the model would suggest the existence of complicated internal structure under the surface of known subatomic degrees of freedom in order to encode large amounts of information about their past evolution. One way or another the beables would be elephantine---either by housing so much additional content, or by never forgetting.

\end{document}